# Title: Human genetic admixture through the lens of population genomics

**Authors:**

Shyamalika Gopalan[1,†], Samuel Patillo Smith[2,3], Katharine Korunes[1], Iman Hamid[1], Sohini Ramachandran[2,3,4]*, Amy Goldberg[1]*

* Contributed equally

† Corresponding author, SG shyamalika.gopalan@duke.edu

**Affiliations**

1 Department of Evolutionary Anthropology, Duke University, Durham NC

2 Center for Computational Molecular Biology, Brown University, Providence RI

3 Department of Ecology, Evolution and Organismal Biology, Brown University, Providence RI

4 Data Science Initiative, Brown University, Providence RI

## Abstract

Over the last fifty years, geneticists have made great strides in understanding how our species' evolutionary history gave rise to current patterns of human genetic diversity classically summarized by Lewontin in his 1972 paper, 'The Apportionment of Human Diversity'. One evolutionary process that requires special attention in both population genetics and statistical genetics is admixture: gene flow between two or more previously separated source populations to form a new admixed population. The admixture process introduces unique patterns of genetic variation within and between populations, which in turn influences the inference of demographic histories, identification genetic targets of selection, and prediction of phenotypes. In this review, we highlight recent studies and methodological advances that have leveraged genomic signatures of admixture to gain insights into human history, natural selection, and complex trait architecture. We also outline some challenges for admixture population genetics, including limitations of applying methods designed for single-ancestry populations to the study of admixed populations.

**Introduction**

In his foundational 1972 study 'The Apportionment of Human Diversity', Richard Lewontin demonstrated that the majority of human genetic diversity at a single locus is contained within, rather than between, populations using polymorphism data from a global sample [1]. The field continues to strive to understand the evolutionary processes that shape this important empirical observation. Notably, genomic data have revealed the extent to which one such process–genetic admixture–has been ubiquitous throughout human history and can shape the apportionment of human genetic diversity in ways different from those predicted by classic population genetic models [2–5]. Here we focus on admixture as a population-level process, whereby gene flow occurs between previously diverged source populations, producing new populations with ancestry from multiple source populations. We discuss how recent research on genetic admixture has extended our understanding of the apportionment of human genetic variation.

Beyond the allele-frequency-based summaries of variation studied by Lewontin (1972), variation in admixed populations can be summarized based on ancestry from source populations. These ancestry patterns may vary between admixed populations formed by the same source populations, between individuals within an admixed population, and across loci within an admixed individual (Figure 1). Population geneticists have long recognized that studying admixed groups provides unique opportunities to learn about evolutionary forces [2,3]. Despite this early interest, inclusion of admixed populations in genetic studies is variable by research goal. Whereas the demographic and selective histories of admixed populations is well-studied, phenotypic and medical studies of admixed populations have lagged behind relative to studies of single-ancestry populations. For example, admixed populations are underrepresented in biobank datasets. The lack of medical genomic samples and the frequent need for admixture-specific methods lead to admixed populations often being excluded from these studies [6–10]. Additionally, in practice, defining admixture in humans is highly context dependent, influenced by societal biases, as well as methodological limits on detecting admixture from genomic data (Box 1).

Many population genetics methods and analytical results are based on assumptions about populations that do not hold under recent admixture. Under a model of isolation, metrics of genomic diversity often have well-defined theoretical expectations with respect to fundamental parameters of the population's evolutionary history. For example, for a single-ancestry population of a given effective population size, linkage disequilibrium and heterozygosity are expected to be inversely correlated with each other, reflecting the relative strength of genetic drift. However, many of these relationships break down with admixture, which changes both linkage structure and allele frequency distributions, potentially confounding standard inference methods. Admixture also complicates the identification of genetic underpinnings of complex traits. For example, genome-wide association (GWAS) studies assume that there are no systematic differences in the genetic variation of the study cohort except in those variants that underlie the phenotype of interest. Yet patterns of global and local ancestry vary widely across individuals within an admixed population (Figure 1). This type of population structure cannot be resolved by simply controlling for genome-wide ancestry, and as a result, GWAS studies of admixed populations may have inflated error rates [7,10–12].

Studying the ancestry patterns of present-day admixed groups has revealed information about the demographic histories of their source populations, including those that are uncommon in unadmixed form today [13,14]. For example, high-resolution genetic maps have been constructed based on the frequency of estimated local ancestry switchpoints (i.e. where local ancestry changes from one source to another along a single chromosome), which contains information about recombination rates along the genome [15,16]. Admixed genomes have also enabled the discovery of variant-trait associations and improved genetic risk prediction models beyond what is possible by studying only single-ancestry populations [10,17–20]. Recent methodological improvements have increased the efficiency and performance of local ancestry calling (i.e. the assignment of genomic segments to their population of origin; Figure 1C) [21–24]. These advances have enabled the use of local ancestry patterns in admixed populations to infer demographic history, adaptation, and the genetic bases of complex traits. When analytical and sampling strategies explicitly account for the admixture process, studies of admixed populations hold tremendous potential for answering fundamental population genetic questions

Here, we consider three inferential problems in which studying patterns of genetic variation produced by admixture is particularly useful: inference of population history, identifying adaptive mutations, and phenotype associations and prediction using admixed genomes. We summarize recent progress in the field, highlight as yet unresolved issues, and outline potential avenues of future research on the genetics of admixed populations. We focus on recent admixture between modern human populations, roughly corresponding to admixed populations founded within the last tens of generations; Witt et al. in this issue consider ancient admixture events with archaic humans and their consequences for human genetic variation [25].

**Estimating genetic diversity and ancestry in admixed populations**

Well before polymorphism data could be generated at a genome-wide scale, several methods of measuring genetic diversity had already been proposed, including heterozygosity and nucleotide diversity [26–28]. By connecting these to theoretical population genetics models, diversity metrics such as those used by Lewontin (1972) can provide insight into the evolutionary forces acting on populations. However, inferring population history from genetic data is highly dependent on how groups are defined, a choice by the researchers (Box 1). Recent admixture complicates the quantification and analysis of genetic diversity, and can therefore affect traditional summaries of diversity in unexpected ways.

In his 1972 paper, Lewontin discusses his choices of a genetic diversity measure at some length, ultimately settling on one that is analogous to heterozygosity and tends to up-weight inter-group differences [1]. Relevant to admixed populations, Lewontin specifically notes that:

> *"a collection of individuals made by pooling two populations ought always to be more diverse than the average of their separate diversities, unless the two populations are identical in composition (pg. 338)"*

In this statement, Lewontin describes expectations of diversity in a set of pooled haplotypes originating from individuals of distinct ancestries, as might result from sampling schemes that combine populations in genetic analysis. This quote gives insight into how admixture may impact the measures of genetic variation that Lewontin considers. These ideas were revisited in a recent study that systematically explores patterns of heterozygosity in admixed populations [29]. The authors analyze a range of simulated admixture scenarios and find that the heterozygosity of an admixed population is at least as high as that of the least heterozygous source population, and under certain conditions, can exceed the heterozygosities of all sources [29]. Thus, the heterozygosity of an admixed population cannot be predicted only from the ancestral sources [29]. $F_{ST}$, central to Lewontin's 1972 analysis, can also be highly informative about the parameters of the admixture process itself, as Boca and Rosenberg demonstrated [30]. These studies demonstrate how traditional measures of genetic diversity can be repurposed to improve our understanding of the admixture process.

Beyond within- and between-population estimates of genetic diversity and ancestry, admixed populations introduce another class of summaries of genetic variation: tracts of the genome *within individuals* that originate from each ancestry source [5,31–33]. In Figure 1, we illustrate three hierarchical categories of genetic ancestry variation in admixed populations from the 1000 Genomes projects from the Americas, who have African, European, and Amerindigenous ancestry [34]. First, given similar continental source ancestries, admixed populations can vary in the average proportions of each source (Figure 1A). Second, individuals within an admixed population may vary in their genome-wide, or 'global', ancestry proportions (Figure 1B). Third, individuals with similar source ancestry contributions and admixture histories may vary by 'local' ancestry across genomic loci (Figure 1C). At each level, these patterns of diversity contain information about admixture and post-admixture processes.

In practice, genetic ancestry is not fully known and is inferred, often using reference panels that are collated to represent the source populations [4,21–24,35–39]. In the following sections, we discuss aspects of human evolution that are commonly inferred from patterns of genetic variation in admixed populations, particularly genetic ancestry. The performance of these methods is predicated on accurate estimates of mean ancestry within a single individual from a given source population, known as global ancestry, or on locus-specific assignment of ancestry within an individual, known as local ancestry.

The quality of ancestry estimates depends on a variety of sampling and evolutionary scenarios [40]. A recent study of the admixed Ashkenazi Jewish population noted that the lack of differentiation between European and Middle Eastern haplotypes made accurate local ancestry inference challenging, reducing their power to infer the parameters of the admixture process [41]. The authors suggest that these issues might be mitigated by the continued development of frameworks that incorporate uncertainty in local ancestry estimates into complex demographic scenarios. Lawson et al. (2018) demonstrate multiple avenues for potential over- or misinterpretation of global ancestry estimates from a commonly used suite of model-based methods based on the Pritchard-Stephens-Donnelly model of mixed membership across latent clusters [40]. For example, multiple qualitatively different evolutionary scenarios produced similar global ancestry estimates in the admixed population, and uneven sample sizes between populations may influence ancestry estimates. Notably, many methods rely on the use of reference panels of modern populations as proxies for the source populations, which may not fully represent the populations that existed at the time of admixture, and have uneven representation globally (Box 1).

**Inferring population history**

The admixture history of a population, such as the timing and source contribution levels, leaves predictable patterns of genetic variation within and between individuals from the admixed population [32,33,42–45]. Empirical genetic analyses can therefore be used to infer the histories that produced observed genetic variation.

Under a simple admixture scenario, the allele frequency of a locus in the admixed population is expected to be the average of the allele frequencies in source populations weights by their contribution levels [46–48]. That is, the admixture contributions from the sources can be estimated from the allele frequencies of the admixed and source populations. Ancestry-based estimation often relies on identifying a subset of loci with particularly large allele frequency differences between populations, known as Ancestry Informative Markers (AIMs) [49,50]. With further developments in genome sequencing and higher coverage of loci across genomes, recent methods often incorporate linkage information or model small allele frequency changes over many loci, producing estimates of genome-wide ancestry proportions (global ancestry), as well as local ancestry along a genome of an admixed individual [4,21–24,35–39]. Mechanistic models of admixture complement empirical studies to improve our intuition of admixture dynamics and interpretation of empirical results [32,44,51–54]. Related model-based inference frameworks have been developed to estimate parameters of population history.

Patterns of global and local ancestry within and between individuals are informative about admixture histories. For example, over time, recombination tends to break up local ancestry tracts; therefore, longer tracts generally indicate more recent contributions from source populations to an admixed population and may be used to infer the timing of admixture [32,44,55–60]. Similarly, as random mating leads to averaging of ancestry proportions, the variance in global ancestry decreases over time as well [33]. Non-ancestry-specific information, such as linkage disequilibrium, can also inform the timing of admixture as populations with differentiated allele frequencies mix.

With two-way admixture, high frequency variants from each source will be strongly correlated with each other in the first-generation admixed population, regardless of their respective locations in the genome. Over time, recombination will erode these correlations to generate a pattern of admixture linkage disequilibrium decay over genomic distance. Several methods leverage these characteristic decay curves to estimate the age of a pulse of admixture [4,61,62], and extensions of these methods infer admixture parameters under models that include continuous gene flow, multiple waves, or assortative mating [63–65].

    Similarly, sociocultural practices that govern mate choice or sex-specific contributions from the source populations will leave signatures in patterns of genetic ancecstry. Individual behaviors such as mating preferences or long-range migration can exhibit ancestry biases, driven by correlations between ancestry and visible traits like skin pigmentation [66–70]. Simple models of admixture often assume individuals mate randomly; however, admixed human populations show evidence of positive assortative mating, with mating pairs often correlated in global ancestry proportion [69,71–73]. Recent methods have sought to test this hypothesis by developing frameworks to infer parental ancestries from phased haplotypes within a single individual [74–76]. Nonrandom mating patterns can bias inference of admixture parameters when not accounted for [64,77]. Additionally, based on the sex-specific inheritance of the X chromosome (where females inherit two copies, one from each parent, while males inherit one X chromosome maternally and their Y chromosome paternally), comparisons of X-chromosomal and autosomal ancestry proportions have be used to infer sex-biased admixture in ancient and modern human populations [52,53,77–83]. These

differences in female and male contribution levels from the sources may be indicative of social interactions between the admixing populations.

Differences in ancestry proportion across the geographic span of a population or populations with shared ancestry components have been used to infer ancestry-biased migration patterns, which may be driven by social cues. For example, ancestry-biased migration, often combined with other mating dynamics, has been proposed as a process shaping regional variation in African ancestry proportions across the USA [67,84,85]. Similarly, temporal changes in ancestry proportion within a population may be caused by time-varying social dynamics. Spear et al. (2020) found a significant increase in Amerindigenous ancestry in Mexican American populations over time [86].

Sufficiently accounting for these complexities presents an exciting challenge. One increasingly popular solution involves simulation-based demographic inference frameworks, such as approximate Bayesian computation and machine learning-based approaches. For example, MetHis is an approximate Bayesian computation-based approach for inference under complex two-way admixture models [51,87]. An advantage of such methods is that they can, in theory, handle arbitrarily complicated admixture scenarios, accommodate any calculable feature of genomic data (such as IBD and ROH), and even conduct summary-statistic-free inference [88]. Continued work to extend these methods will enable disentangling the myriad historical, evolutionary, and socio-cultural factors contributing to admixture processes.

Studying the genomes of admixed populations can also provide insight into the genetic origins and demographic histories of their founding populations, particularly for some source ancestries that are no longer commonly represented by an extant single-ancestry population [13,14]. An increasingly popular approach is to first parse admixed genomes by local ancestry and before applying classic single-population methods on each source separately. This is exemplified by the ancestry-specific PCA (ASPCA) method, which performs PCA separately to each contributing source ancestry, as identified by local ancestry inference methods. This approach has revealed previously unappreciated variation in the European and American ancestry sources of admixed Latinos across across Mexico [89], the Caribbean [90], and South America [91].

Local ancestry inference can also be used to unravel source-specific historical population size dynamics. The process of admixture often involves bottlenecks at the time of founding, which Browning et al. (2018) demonstrated can be inferred using ancestry-specific identity by descent (IBD) [92]. This approach combines estimates of local ancestry and IBD for admixed groups to estimate the past effective population sizes of each of the source ancestries. They find population-specific size changes within continental ancestries.

Moving forward, combining ancestry-based inference with other summaries of genetic variation may help elucidate these complex and dynamic population histories. For example, patterns of homozygosity and IBD are shaped by the relationships between mating pairs and can contain additional information about historical and sociocultural processes [67,90,93,94]. However, we lack theoretical expectations for the distributions of runs of homozygosity (ROH) and IBD segments after admixture, which may break up homozygosity while also involving major changes in population size and structure. Recent empirical explorations in admixed populations suggest that, in particular, RoH in admixed populations reflect both contributions from source populations and post-admixture population dynamics, with implications for the distribution of deleterious variation.

**Detecting selection**

Adaptation to biotic and abiotic environments leaves signatures in patterns of human genetic variation that can be used to infer the selection history and identify adaptive loci [95–98]. However, admixture can confound and obscure features of genetic data that are classically interpreted as signatures of selection [99–102]. Additionally, admixture may introduce novel selective pressures through large geographic movement of people. Under certain scenarios, selection may indeed be easier to detect in admixed populations than single-ancestry populations with the added information from ancestry patterns [17,103,104]. That is, inferring selection from admixed genomes poses unique challenges, but also opportunities for new insights into human adaptation.

As described earlier, admixed populations are often considered as a linear combination of their sources. Under this model, the expected allele frequency of a locus in the admixed population is an average of the allele frequencies in each source population at that locus weighted by the contributions of each source population to the admixed population. Loci that dramatically differ from this expectation are candidates for loci under selection (reviewed in Adams & Ward 1973, and Chakraborty 1986).

Outlier methods have been used to detect selection with a variety of summary statistics in single-ancestry populations, including early work by Lewontin and Krakauer (1973) [105], and more recently, IBD or ROH. However, demography can change the distribution of these statistics across the genome, leading to spurious results or complicating interpretation of these outlier methods [106–109]. When using methods not specifically developed for admixed populations, admixture can lead to both increased false positive rates and decreased power to detect both pre-admixture selection (i.e. selection that happened in the source populations) and post-admixture selection [100]. This may partly explain why previous studies using African-American populations as a proxy for African populations suggested that Africans have experienced significantly less selection than other human groups [110].

Recent methods often leverage ancestry information to detect post-admixture adaptation, independently based on ancestry distributions or in combination with other classic summary statistics [103,104,111–115]. When selective pressures are shared between admixed populations and one of their sources, admixture-mediated adaptation may occur through contributions of an adaptive allele from that source population. This may be a particularly rapid mode of adaptation because the allele may be introduced into the admixed population at intermediate to high frequency (proportional to the admixture contribution from that source), decreasing stochastic loss. If the adaptive allele is common in one source population but rare in the other(s), then as that allele rises in frequency in the admixed population, so will the corresponding local ancestry. This observation has led to a common method to detect post-admixture selection: scanning for outliers in local ancestry compared to genome-wide ancestry.

Empirical studies have identified numerous candidate regions under selection post-admixture using ancestry outlier methods [111,116–118]; however this approach has several limitations. The distribution of local ancestry within a population is influenced by a complex interplay of selective and demographic histories, and current theoretical understanding is limited, making the choice of cutoff for identifying outliers somewhat arbitrary [119]. More fundamentally, an ancestry-outlier approach is only suitable in situations where the allele frequencies in the source populations differ substantially, which couples allele frequency changes with a single source's ancestry. Figure 2 demonstrates this coupling; the proportion of simulations in which the adaptive locus is an outlier increases with increasing $F_{ST}$, a measure of allele frequency difference between the sources. We simulated admixture with equal contributions from the sources followed by 12 generations of strong selection (*s*=0.05) at the adaptive locus. Additionally, the power for outlier approaches to localize adaptive loci depends on the length of the admixture tract containing the loci, and therefore the selection history. Finally, while useful for identifying adaptive loci, these methods must be combined with other information or simulations to infer parameters of the history such as the strength or timing of selection.

Ongoing work extends initial implementations of ancestry-outlier approaches to study post-admixture selection, and often uses simulations to improve interpretation and test power [103,104,112,113,120,121]. These methods have recovered classic examples of selected loci from the genomes of admixed populations, and inferred the timing, strength, and repeatability of selection under different scenarios. For example, in Hamid et al. (2021), we found signatures of adaptation to malaria via the *DARC* gene in the admixed population of Cabo Verde based on long, high frequency African ancestry tracts. We further used simulation-based inference to infer the strength of selection. This study's findings reinforced others that have identified post-admixture selection pressure to retain African ancestry at *DARC*, a known malaria susceptibility locus, in multiple admixed populations [103,118,122–124]. It also provides an example of combining ancestry-specific summary statistics with simulations to both localize selection and infer parameters of the selection history.

While these recent studies using empirically-driven summary statistics have proven informative in certain scenarios, the field is only just beginning to develop expectations of the distributions of ancestry under models of selection with admixture. Indeed, recent work has suggested perhaps unexpected relationships between ancestry tract lengths, allele frequencies, and selection history, emphasizing the need for additional theory [113].

**Understanding complex trait architecture and predicting genetic traits**

For decades, human genetics research has aspired to make personalized medical therapies a reality; while progress has been made on the genetic prediction of traits in recent years, its potential benefits are still only readily applicable to individuals of European ancestry [125]. Over the past decade, genome-wide association (GWAS) studies have become a standard framework for studying the genetic basis of complex traits, in which variants across the genome are tested individually for statistical association with a phenotype of interest. GWAS studies have also formed the statistical foundation for polygenic scores (PGS), in which complex quantitative traits are predicted (e.g. height or cholesterol level) under Fisher's infinitesimal model from the sum of an individual's observed genotypes weighted by GWAS-inferred effect sizes (Figure 3).

However, admixture introduces complex population structure and linkage blocks that, if unaccounted for, can identify false positive associations. Recent research has shown that variant effect sizes estimated from GWAS studies tend to be ancestry- or even study-specific (Figure 3) [126–129]. This severely limits the transferability of PGS across different human ancestries, and in practice, results in poor trait prediction for non-European individuals or for any individuals who were not part of the discovery GWAS [126,128]. Underscoring this point, recent studies have shown that when calculating a European ancestry-based PRS for admixed individuals, predictive accuracy increases with the individual's genome-wide proportion of European ancestry [86,130,131]. Recent research has developed frameworks to improve performance of PGS in individuals from admixed populations, such as including local ancestry-based principal components to correct for heterogeneous patterns of population structure along the genome or subdividing the cohort by genome-wide ancestry and taking a meta-analysis approach [125].

PRS accuracy could be improved with more comprehensive sequencing of cohorts of non-European ancestry [8,127,132], but must be coupled with new methods tailored to admixed populations and their linkage disequilibrium patterns and allele frequency variation due to unique population histories and selection pressures (see also Fish et al. 2018 [133]. Admixture can increase the likelihood of uncommon genotype-genotype interactions, and can occur in tandem with migration events that can generate novel genotype-environment interactions [134]. Furthermore, source ancestry contributions to admixed populations and their dynamics within admixed populations can change over time, leading to temporal variation in effect size estimates [86]. All of these factors can contribute to a loss of predictive power in admixed individuals even when accounting for local ancestry and using high-quality effect size estimates for all source ancestries [135,136].

Admixture can also provide unique information and significantly more statistical power for characterizing the genetic basis of complex traits [20,86]. An early example of this was admixture mapping, which tests for associations between a trait and local ancestry under the assumption that one of the source populations has elevated disease risk relative to the others. This approach has been successful in identifying ancestry-specific effects on disease susceptibility in admixed cohorts [17–19,137–139]. More recently, methods using local ancestry information have yielded additional insights into the genetic drivers of complex and molecular phenotypes, which can be confounded by admixture even when they are not strongly stratified by ancestry [10,140]. By incorporating local ancestry calls from African-Americans in their analysis, Zhong et al. (2019) show an improvement in the accuracy of eQTL mapping and heritability estimation in admixed populations [140]. Similarly, a recent study found that explicitly accounting for local ancestry in GWAS studies can result in higher power to identify and fine-map causal variants [10].

Increasingly, research suggests that by excluding individuals from admixed and minority populations, geneticists are discarding a rich source of genomic information [20]. It is therefore critical for studies of quantitative trait architecture to move forward with addressing inequities in sampling while also developing statistical methods that appropriately model the dynamic evolution and unique characteristics of admixed populations' genomes.

**Conclusions**

Though it was not the focus of his paper, Lewontin (1972) acknowledged the role of admixture in distributions of genetic variation, and included admixed populations in his analyses (see also Box 1). In the intervening fifty years, population genetics research has continued to shed light on the importance of admixture processes in shaping genetic variation and complex trait architectures. In certain scenarios, studying admixed populations may provide insight into general human evolutionary processes and history beyond admixture itself because of the added information from ancestry-based statistics.

Multiple future directions in research on admixture will extend our understanding of human evolution and the distribution of human genetic variation. Firstly, whereas certain demographic histories are well understood theoretically, there is an urgent need for theory regarding how natural selection interacts with admixed population histories. Figure 2, as well as multiple recent studies [30,105,106,108,141,142], show that common summary statistics to detect selection in admixed populations have variable power and often unclear interpretations; researchers should be cautious when applying outlier scans for genomic targets of selection to admixed populations. Secondly, the study of admixed populations is often based on contrasting genetic variation in admixed populations against that of reference populations for source ancestries, even if accurate references are not available. Reference-free methods for ancestry assignment as well as inference would move beyond this challenge, perhaps reducing bias [5], as well as more inclusive sampling. Finally, multiple recent studies have focused on methods for predicting quantitative traits in admixed populations [86,127,130,131,140], and there are opportunities to broadly study how admixture linkage disequilibrium specifically influences the identification of genetic associations and the ability to predict quantitative traits.

**Box 1 - Lewontin (1972) and admixed populations**

Although Lewontin (1972) used multiple admixed populations in his analyses, his goal — to argue definitively against the use of genetic loci to reinscribe racist divisions among human groups (see also Novembre, and Shen and Feldman in this volume) — did not require special consideration of genetic processes underlying and genetic variation in admixed populations. When presenting the Lewontin (1972) result (see also Novembre in this volume), discussing admixed populations may prove helpful to gain new insight into the lasting legacy of Lewontin (1972).

While research by Lewontin and many others has shown that genetic differences between populations comprise a small proportion of human genetic diversity at a given locus [1,2,35,143,144], it is important to note that discussions of admixture are, implicitly or explicitly, predicated on the idea that meaningful genetic differences exist between discretized human groups. Lewontin's entire body of work, and in particular his 1972 paper, pushed back on these ideas (see Shen and Feldman in this issue) [1,145,146]. In practice, the term "admixed" can vary to encompass a range of spatial and temporal processes of gene flow between previously isolated groups. Extensive gene flow between groups is a hallmark of recent human evolution; essentially all extant populations descend from a combination of multiple distinct ancestries. In population genetics, however, the term 'admixed' is used to refer to a relatively narrow subset of human groups that trace their ancestry to recent combinations of two or more previously genetically isolated groups [3–5,32,33]. However, there are no strict criteria that determine which populations should be considered admixed from a genetic perspective, making classification a highly subjective process [33,42]. Additionally, while the effects of genetic admixture can be observed in individual genomes, it is conceptualized as a demographic process that acts on populations. For example, for a recent two-way admixture pulse between populations A and B, high variance in these ancestry components across individuals is expected; under neutrality, there will be individuals in the admixed population that derive 0%, and those that derive 100%, of their ancestry from population A (Figure 1) [33,42]. That is, finding a representative genome of an admixed population is an ill-defined goal. As demonstrated in the bottom panel of Figure 1, no individual locus in a genome from an admixed population will enable inference of the demographic history of the admixed group. While Edwards (2003) [147] response to Lewontin (1972) emphasized the importance of multilocus genetic diversity in enabling ancestry inference from genetic data, genomes from the same admixed population can have substantial variance in how ancestral diversity is distributed, underscoring Lewontin's argument that genetic data does not suggest a classification of human groups. In fact, the identification of ancestry-informative markers [49,50] for admixed groups is based on identification of loci with high population-specific differentiation, a contrast to the general pattern observed by Lewontin (1972).

    Methodological issues and patterns of genetic diversity are not the only factors shaping our understanding of genetic admixture; science and society's preoccupation with classifying human groups using largely superficial characteristics also plays a role. Lewontin aptly pointed this out in his 1972 paper:

*"The erection of racial classification in man based upon certain manifest morphological traits gives tremendous emphasis to those characters to which human perceptions are most finely tuned (nose, lip and eye shapes, skin color, hair form and quantity), precisely because they are the characters that men ordinarily use to distinguish individuals. Men will then be keenly aware of group differences in such characters and will place strong emphasis on their importance in classification. (pg. 382)"*

Geneticists and anthropologists have long wrestled with the various field-specific and lay definitions of 'population', 'ancestry', 'ethnicity', and 'race', which interact and intersect with each other in complicated ways [31,148–151]. In discussions of human genetic admixture, it becomes especially important to emphasize that these categories do not map on to each other one-to-one, and that race and ethnicity, in particular, are social constructs. The variation in ancestry within individuals in admixed populations, shown in Figure 1, illustrates this and can be an effective tool in teaching the difference between genetic ancestry, phenotypes, and self-identified race and/or ethnicity.

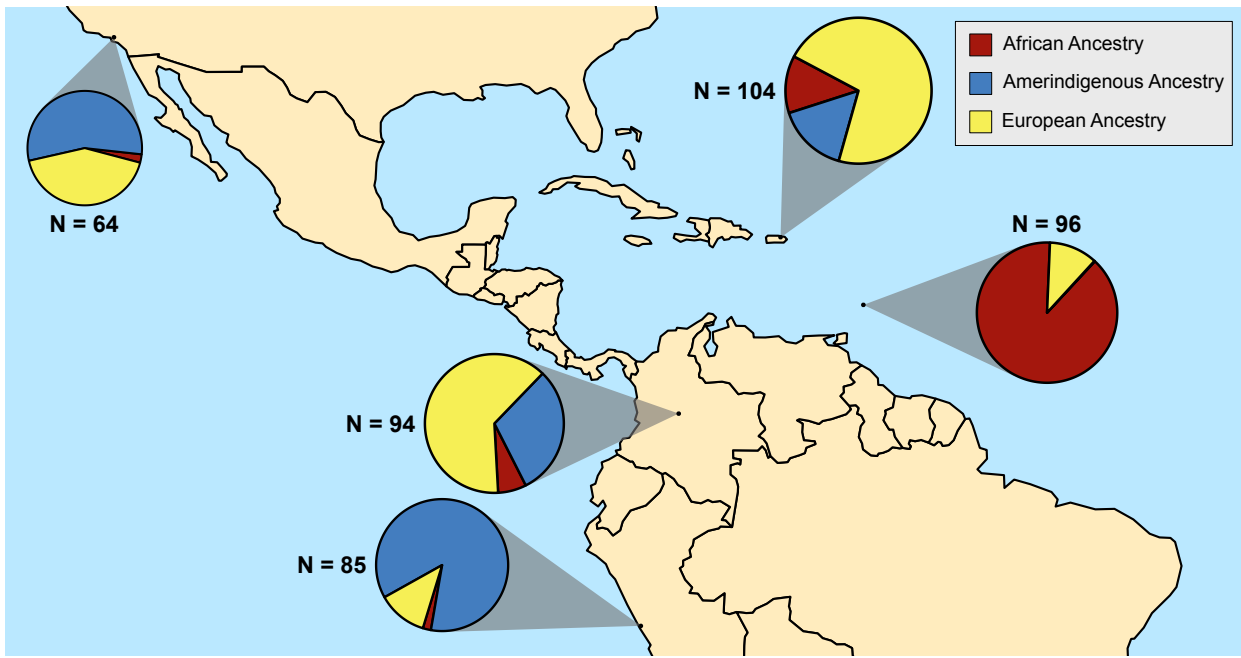
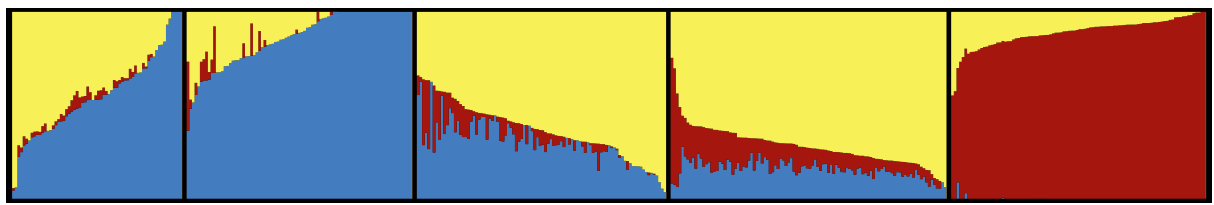
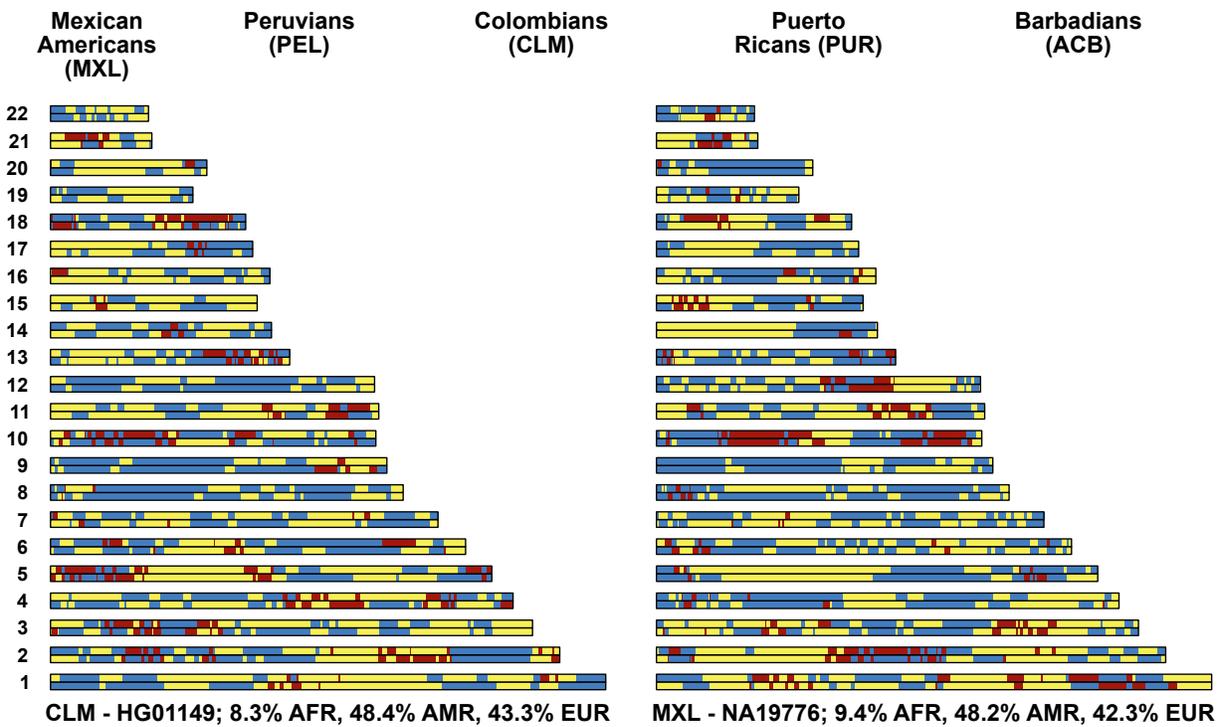

**CLM - HG01149; 8.3% AFR, 48.4% AMR, 43.3% EUR**   **MXL - NA19776; 9.4% AFR, 48.2% AMR, 42.3% EUR**

**Figure 1 - Ancestry in admixed populations varies at multiple genetic scales, with variance among individuals and within individual genomes.** We plot examples of global and local ancestry inferred from 1000 Genomes Project populations from the Americas and Caribbean (MXL, PEL, CLM, PUR, and ACB), using West African and European populations as references for the sources (ESN, GWD, MSL, YRI, IBS, and TSI). Global ancestry was estimated in ADMIXTURE; barplots made using pong (barplots, middle panel) [152]. Local ancestry inferred for two example individuals with similar global ancestry proportions (HG01149 from CLM and NA19776 from MXL), using RFmix. We filtered related individuals, SNP missingness (> 5%), minor allele frequency (< 1%), Hardy-Weinberg disequilibrium (p-value < 0.000001), and linkage disequilibrium (using the PLINK command --indep-pairwise 50 10 0.1).

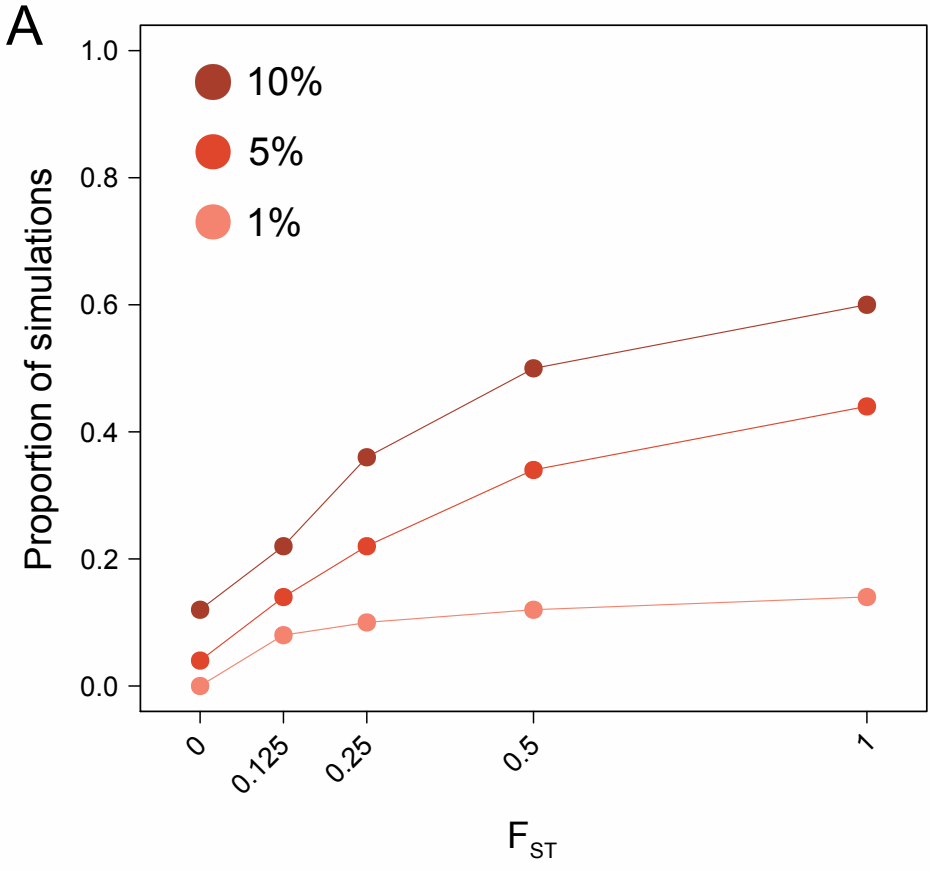

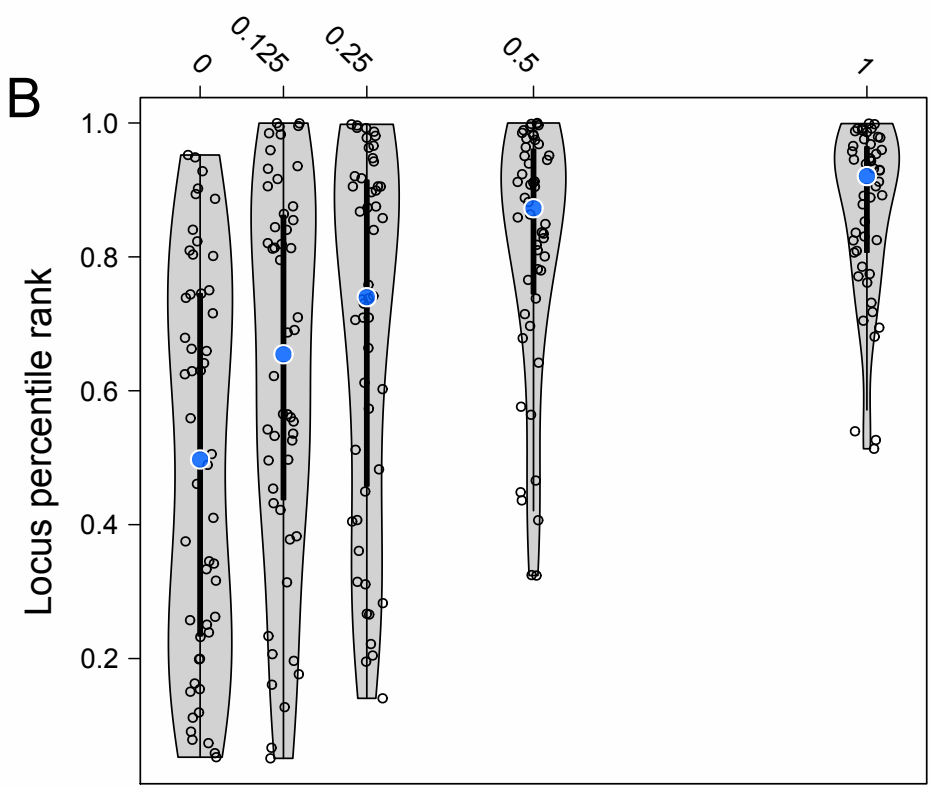

**Figure 2 - Ancestry outlier tests for post-admixture selection are underpowered when source differentiation is low.** We examine how $F_{ST}$ between two source populations at a selected locus affects the power of a local ancestry outlier approach to detect selection. Whole-genome simulations were conducted in SLiM [153]. We simulated 50 sets of 10,000 individuals under a two-way admixture model with equal contributions from the sources with Population A contributing an allele that is under strong selection (s = 0.05) in the admixed population for 12 generations. For increasing values of $F_{ST}$ along the x axis, we plot (A) the proportion of simulations in which the selected locus would be classified as an 'outlier' in local ancestry frequency from Population A for multiple genome-wide thresholds, and (B) the rank of the selected locus among all loci genome-wide for ancestry from Population A. Even with relatively strong selection and complete differentiation between source ancestries (i.e. $F_{ST}$ = 1) at the selected locus, it frequently failed to appear as a Population A ancestry outlier. We note, however, that we only allowed for 12 generations of selection; if we allowed the simulation to proceed for longer, selection would continue favoring Population A ancestry at that locus, driving it past a given outlier threshold. Similarly, the rank of the selected locus for Population A ancestry increases with increasing differentiation between source populations at the locus. We simulated 6 diploid individuals per source population, and calculated the allele frequency combinations that would produce each desired value of $F_{ST}$ where Population A's frequency was equal to or higher than Population B's. From these, we randomly chose a starting allele frequency combination for the source populations for each of the 50 simulations.

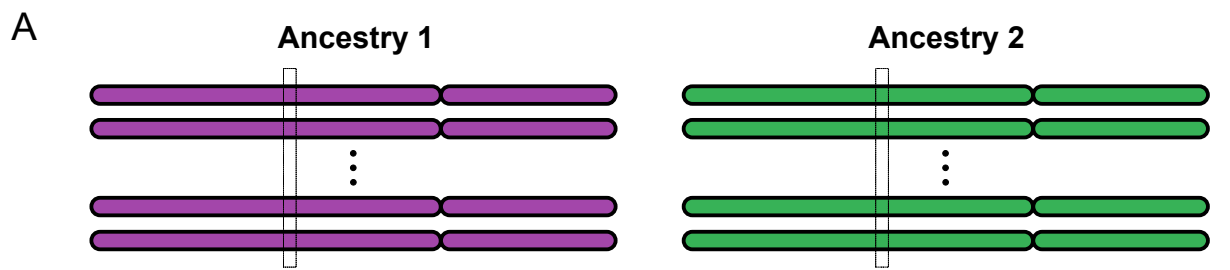
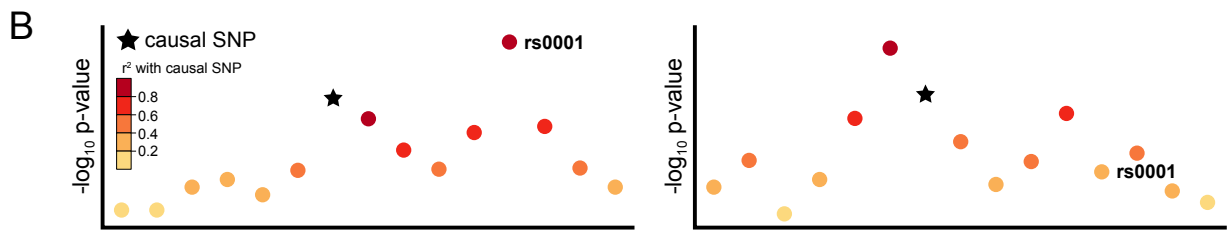
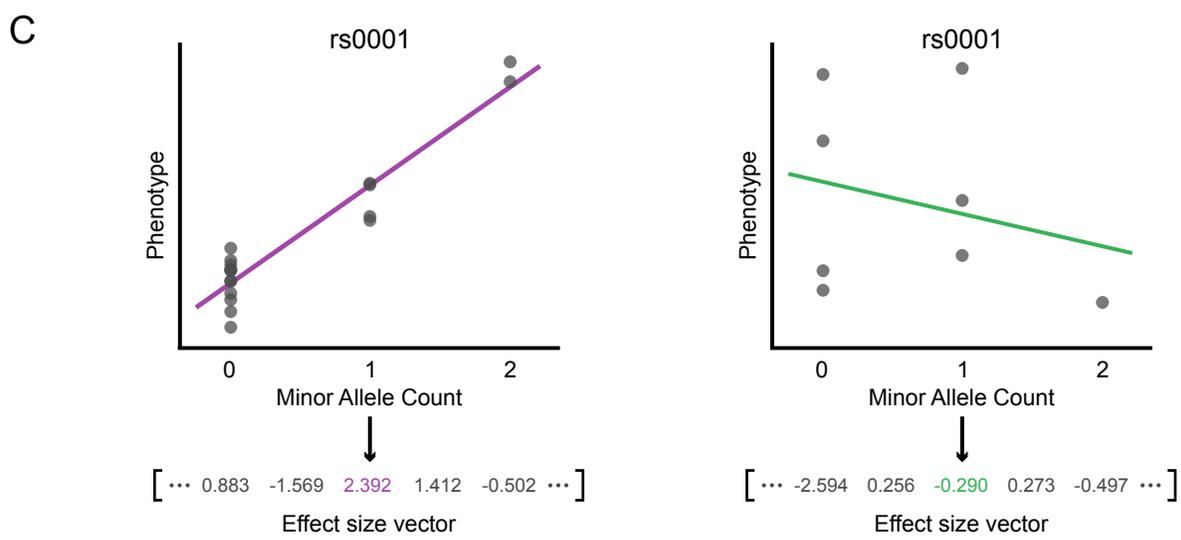
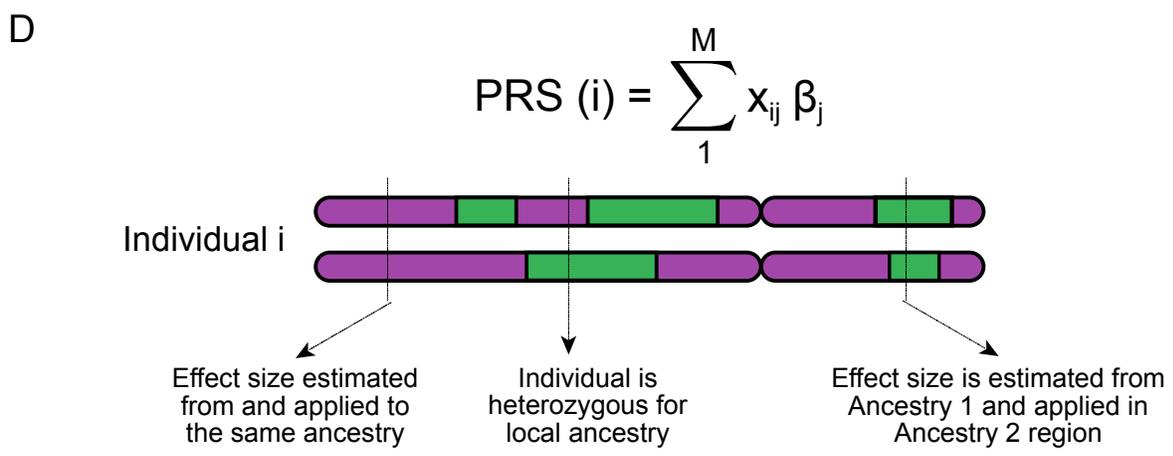

**Figure 3 - Effect size heterogeneity in ancestral populations impedes accurate phenotype prediction in admixed individuals.** We show a scenario in which two single-ancestry populations (A) share a genetic architecture for a trait, yet have different inferred effected sizes and directions inferred through GWAS. The causal SNP (star) is plotted along with 14 other nearby variants, colored by their correlation with the causal variant (B). In Ancestry 1, the SNP most strongly associated with the phenotype is 'rs0001'. This SNP tags the causal SNP, which is frequently unknown in practice. However, rs0001 is poorly associated with the causal SNP, and the phenotype, in Ancestry 2. The effect size of a particular SNP on the phenotype is estimated by regressing an individuals' phenotype value onto their minor allele count (C), showing differences between ancestries despite the shared genetic architecture for the true causal SNP. Ancestry-associated patterns of linkage disequilibrium, allele frequency, and sample sizes, as well as SNP ascertainment bias, can drive variation in tag SNPs and effect size estimates. An individuals' PRS is calculated by summing the effect sizes ($\beta$) at each of $M$ SNPs, weighted by their dosage ($x$) of the minor allele (D). How best to apply this framework to admixed populations is an area of ongoing research.

151. Van Arsdale AP. 2019 Population demography, ancestry, and the biological concept of race. Annu. Rev. Anthropol.. 48, 227–241. (doi:10.1146/annurev-anthro-102218-011154)

152. Behr AA, Liu KZ, Liu-Fang G, Nakka P, Ramachandran S. 2016 pong: fast analysis and visualization of latent clusters in population genetic data. Bioinformatics 32, 2817–2823.

153. Haller BC, Messer PW. 2019 SLiM 3: Forward genetic simulations beyond the Wright-Fisher model. Mol. Biol. Evol. 36, 632–637.
46